\documentclass[sigconf]{acmart}

\AtBeginDocument{%
  }

\copyrightyear{2026}
\acmYear{2026}
\setcopyright{cc}
\setcctype{by}
\acmConference[BuildSys '26]{The 13th ACM International Conference on Systems for Energy-Efficient Buildings, Cities, and Transportation}{June 22--25, 2026}{Banff, AB, Canada}
\acmBooktitle{The 13th ACM International Conference on Systems for Energy-Efficient Buildings, Cities, and Transportation (BuildSys '26), June 22--25, 2026, Banff, AB, Canada}
\acmDOI{10.1145/3744256.3812578}
\acmISBN{979-8-4007-2012-3/2026/06}

\author{Ava Mohammadi}
\orcid{https://orcid.org/0009-0004-9596-4482}
\affiliation{%
  \institution{Eindhoven University of Technology}
  \city{Eindhoven}
  \country{Netherlands}
}
\email{a.mohammadi@tue.nl}

\author{Rick Kramer}
\affiliation{%
  \institution{Eindhoven University of Technology}
  \city{Eindhoven}
  \country{Netherlands}
}
\email{R.P.kramer@tue.nl}

\author{Zoltan Nagy}
\affiliation{%
  \institution{Eindhoven University of Technology}
  \city{Eindhoven}
  \country{Netherlands}}
\email{z.nagy@tue.nl}

\title{Coordination Architecture Shapes Continuous Demand Response Outcomes in Building Districts}

\date{December 2025}

\usepackage{subcaption}
\usepackage{booktabs}
\usepackage{float}
\usepackage{graphicx} 
\usepackage{siunitx}  
\usepackage{multirow}
\usepackage{placeins}

\usepackage{amssymb}

\usepackage{todonotes}

\begin{document}

\begin{abstract}
Grid-integrated building districts must provide energy flexibility while preserving 
occupant comfort and equitable distribution of control burden. We study how 
coordination architecture influences the ability of building clusters to track 
aggregated load profiles, comparing four paradigms: centralized model predictive 
control (MPC), decentralized independent reinforcement learning (SAC), 
centralized-training-decentralized-execution multi-agent RL (MAPPO), and a hybrid 
MPC--SAC controller that separates district-level battery optimization from 
building-level HVAC regulation. A rule-based controller serves as a baseline. 
We evaluate a 25-building residential district across three metrics: aggregate load tracking, thermal comfort, and spatial variability of control actions. We find that architecture choice determines the trade-off structure. Centralized MPC achieves low tracking bias (8.8\% NMBE) but concentrates actuation on a subset of buildings, causing elevated comfort violations (24.8\% exceedance) and spatial imbalance. Decentralized RL distributes control effort more evenly but fails to sustain accurate tracking. The hybrid architecture achieves the best balance: accurate tracking (4.8\% NMBE), moderate comfort impact (16.8\% exceedance), and the lowest spatial variability. These findings demonstrate that architecture choice determines the trade-off structure between tracking and comfort.

\end{abstract}



\maketitle

\section{Introduction}
Buildings are increasingly recognized as active participants in power system operation, due to the growing penetration of controllable loads, on-site generation, and energy storage systems \cite{li_ten_2022}. HVAC systems, heat pumps, and thermal and electrical storage can provide temporal flexibility. However, this flexibility becomes valuable only when coordinated across many buildings, i.e., when individual responses aggregate into system-relevant capacity, e.g., for load shifting and congestion mitigation~\cite{nweye_merlin_2023}. Thus, the meaning of energy flexibility has shifted from building-level, event-based, load modulation to the ability of aggregated building loads to track grid-requested power trajectories under dynamic signals \cite{mathieu_new_2024}. This can be achieved through energy flexibility contracting, in which building communities commit to follow predefined load modification profiles specified by an aggregator or system operator \cite{el_geneidy_contracted_2020}.

Successful reference load tracking depends on the underlying coordination structure and may induce rebound effects, and uneven allocation of control actions~\cite{nweye_merlin_2023}.  Coordination is the harmonization of individual control actions toward common system-level objectives~\cite{hu_neighborhood-level_2021}, and it determines how flexibility can be mobilized and constrained at the community level~\cite{li_ten_2022}. Control architectures can be categorized into centralized vs. decentralized and  hierarchical vs. fully distributed~\cite{hu_neighborhood-level_2021}. The choice of the control architecture impacts the aggregated load behavior and shapes how aggregate objectives are realized through building-level control.

Model-based control, particularly centralized and distributed model predictive control (MPC), are widely used to realize such coordination while handling comfort and operational constraints~\cite{drgona_all_2020}. Centralized MPC control maximizes tracking accuracy at high computational and privacy cost, while decentralized MPC control sacrifices accuracy for scalability \cite{lefebure_distributed_2022}. The distributed MPC approach can achieve near-centralized MPC performance with limited information exchange and experimental validation~\cite{lefebure_distributed_2022}. Centralized MPC coordination results in more reliable tracking, whereas naive decentralization leads to higher pre-peaks and reduced efficiency~\cite{el_geneidy_contracted_2020}. 

In contrast, learning-based controllers have gained attention for distributed energy management. A field deployment applied a model-free reinforcement learning (RL) controller to 13 buildings in a district heating network, demonstrating peak shaving and energy reduction while preserving thermal comfort~\cite{moshari_real-world_2026}. In simulation, a fully decentralized multi-agent RL (MARL) approach has been demonstrated for coordinated energy management in interconnected buildings and microgrids, where each agent observes only local states and aggregate-level signals \cite{zhang_multi-agent_2023}. While this approach enables scalable coordination, it exhibits slow convergence and sensitivity to non-stationarity caused by simultaneous learning agents. It has also been shown that decentralized MARL in large residential districts can reduce peak demand and operating costs, at the expense of reduced tracking accuracy and increased training instability as the number of buildings increases~\cite{savino_scalable_2025}. Centralized training with decentralized execution (CTDE) remedies this as the agents are allowed to exploit global information during training, while retaining local autonomy during deployment~\cite{charbonnier_centralised_2025}. 
\begin{table}[t]
\centering
\caption{Summary of notation used in the methodology.}
\label{tab:notation}
\footnotesize
\setlength{\tabcolsep}{5pt}
\begin{tabular}{cl}
\toprule
Symbol & Description \\
\midrule
$i$ & Building index, $i \in \{1,\dots,N\}$ \\
$k$ & Discrete time index \\
$N$ & Number of buildings in the district \\

$o_k^i$ & Observation vector of building $i$ at time $k$ \\
$a_k^i$ & Action vector of building $i$ at time $k$ \\
$a_{k,\text{hvac}}^i$ & HVAC control action of building $i$ \\
$a_{k,\text{batt}}^i$ & Battery (BESS) control action of building $i$ \\

$u_k^i$ & HVAC actuation fraction (MPC decision variable) \\

$y_k^i$ & Net electricity consumption of building $i$ \\
$y_k$ & Aggregated district electricity consumption \\
$r_k$ & Aggregator reference load signal \\
$\delta$ & Threshold separating quadratic and linear regimes in the Huber loss \\

$T_k^i$ & Indoor temperature of building $i$ \\
$T_k^{\text{out}}$ & Outdoor air temperature \\
$v_k^i$ & Instantaneous comfort violation of building $i$ \\

$P_i^{\text{HVAC}}$ & Nominal HVAC power of building $i$ \\
$p_k$ & Aggregated BESS charging/discharging power (MPC) \\
$p_{i,k}^{\text{batt}}$ & BESS charging/discharging power of building $i$ at time step $k$ \\
$\mathrm{SOC}_k^i$ & Battery state of charge of building $i$ at time step $k$ \\
$E_{\text{cap}}^i$ & Battery energy capacity of building $i$ \\

$s_k$ & Tracking slack variable in MPC \\
$s_{k,\text{lo}}^i,\; s_{k,\text{hi}}^i$ & Lower and upper comfort slack variables \\

$H$ & MPC prediction horizon \\
$\Delta t$ & Simulation time step (hours) \\

\bottomrule
\end{tabular}
\end{table}

Community-scale building control is an inherently multi-agent problem with key challenges such as non-stationarity, constraint handling, safety, and privacy-preserving coordination~\cite{nagy_ten_2023-1}. Thus, hybrid MPC--RL architectures are suggested as a practical pathway to combine physical feasibility and constraint satisfaction with the adaptability of learning-based control in communities~\cite{nagy_ten_2023-1,hajialigol_hierarchical_2026}, reproducing their success in single-building energy management~\cite{arroyo_reinforced_2022}. 



Demand-side flexibility, defined as ``the capability of any active customer to react to external signals and adjust their energy generation and consumption in a dynamic time-dependent way, individually as well as through aggregation''~\cite{smartenergy2022}, is shifting from event-based curtailment toward continuous, reliable tracking of grid-requested power trajectories~\cite{mathieu_new_2024}. This shift demands active coordination across buildings, yet coordination can produce unequal allocation of comfort degradation and control effort across participants~\cite{soares_review_2024,nagy_ten_2023}. Recent work established that the choice of coordination structure is a fundamental design decision with context-dependent consequences for system outcomes~\cite{charbonnier_coordination_2022}, yet empirical evidence quantifying how these architectural choices shape the joint outcomes of tracking and comfort in building districts remains limited. While centralized MPC, decentralized RL, and hybrid architectures have each been studied for district-level control, existing studies typically assess them on aggregate district-level performance without examining how tracking accuracy and comfort outcomes are jointly shaped by the coordination structure, or whether they are achieved uniformly across the building population.


Thus, this paper addresses the following research question: \textit{how does coordination architecture shape the trade-off between district-level tracking accuracy and building-level thermal comfort under continuous flexibility provision?} We evaluate four control architectures: centralized MPC, decentralized I-SAC, CTDE-based MAPPO, and a hybrid MPC--SAC controller, on a 25-building residential district, assessing tracking fidelity, comfort preservation, and the spatial distribution of control actions as a diagnostic indicator of how each architecture allocates effort across buildings. We show that architecture choice determines the trade-off \textit{structure}: centralized optimization achieves superior tracking at the cost of comfort, while hierarchical decomposition matches control scope to objective scale, achieving the best balance between tracking and comfort.

The next section introduces the four control architectures and evaluation metrics. In Section~\ref{s:results} we present our results. In Section~\ref{s:discussion} we discuss our findings, and we conclude the paper in Section~\ref{s:conclusion}.

\section{Methodology}
Table~\ref{tab:notation} summarizes the notation used in the paper. 

\subsection{Problem Formulation}

\begin{figure}[t]
    \centering
    \includegraphics[trim=0.25cm 0.25cm 0.25cm 0.25cm,clip, width=\columnwidth]{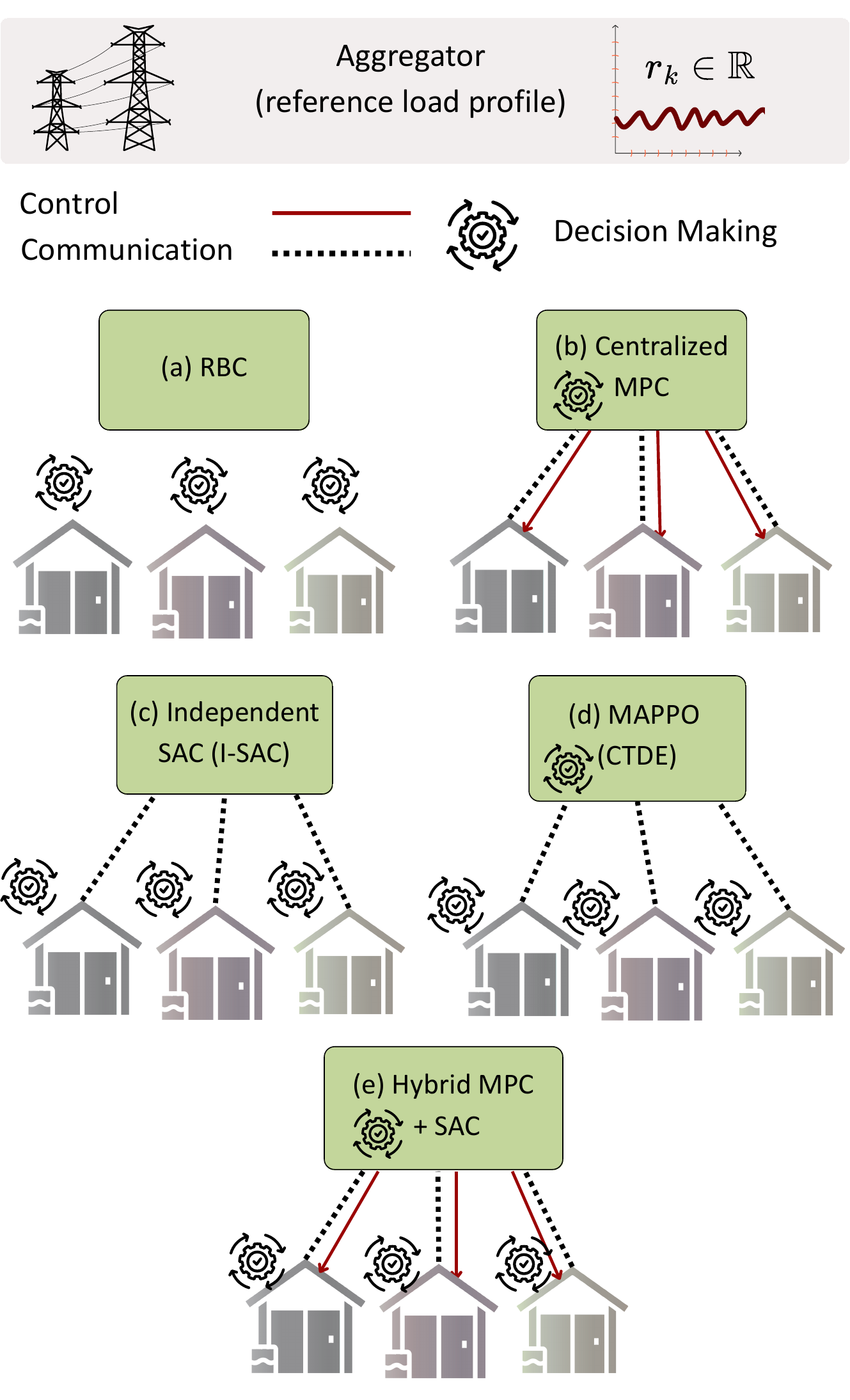}
    \caption{Coordination architectures in this study. The aggregator broadcasts the grid reference signal $r_k$, to (a) the baseline rule based controller, (b) centralized MPC, (c) decentralized I-SAC, (d) centralized-training-decentralized-execution MAPPO, and (e) hybrid MPC--SAC controllers, which generate HVAC and BESS control actions. }

    \label{fig:methodology}
\end{figure}

We consider a district of $N$ buildings interacting with the power grid over discrete time steps $k$. The aggregated district load is
\begin{equation}
y_k = \sum_{i=1}^{N} y_k^i,
\end{equation}
where $y_k^i$ is the net electricity consumption of building $i$.

The district is required to track a reference signal $r_k$ specified by an aggregator. In this study, $r_k$ is defined as a constant daily load profile equal to the average baseline district consumption,
\begin{equation}
r_k = \frac{1}{K}\sum_{j=1}^{K} y_j^{\text{base}}, \quad \forall k,
\end{equation}
where $y_j^{\text{base}}$ denotes the uncontrolled (baseline) district load and $K$ is the episode length in time steps. This formulation represents long-term contractual flexibility commitments rather than short-term event-based demand response. The aggregator broadcasts the district-level reference signal $r_k$ to all buildings; decentralized controllers receive $r_k$ together with local observations $o_k^i$, while centralized controllers additionally access global system states.

The control objective is to minimize the deviation between $y_k$ and $r_k$ while maintaining indoor thermal comfort across all buildings. To achieve this objective, we consider the following coordination architectures (see Figure~\ref{fig:methodology}):  
(a) the baseline rule-based controller (RBC) that combines time-of-use battery scheduling with thermostat-based HVAC control to preserve thermal comfort without learning or system models ;  
(b) a centralized model-based control (MPC), where a single optimizer jointly computes HVAC and battery actions using the full district state;  
(c) a fully decentralized learning-based control using independent Soft Actor--Critic (I-SAC), where each building acts based on local observations;  
(d) a centralized training with decentralized execution using MAPPO; and  
(e) a hierarchical hybrid architecture in which centralized MPC enforces district-level objectives through battery control while decentralized SAC agents regulate building-level HVAC dynamics.

\subsection{Control Architectures}

Tables~\ref{tab:info_access} and~\ref{tab:coordination} summarize the information structure and control responsibilities of our coordination architectures. 

\begin{table}[t]
\centering
\caption{Information available to each controller at runtime (indicated by checkmarks).}
\label{tab:info_access}
\footnotesize
\setlength{\tabcolsep}{4pt}
\resizebox{\columnwidth}{!}{%
\begin{tabular}{lccccc}
\toprule
Variable & RBC & MPC & I-SAC & MAPPO (exec) & Hybrid \\
\midrule
Aggregated load $y_k$                  &  & \checkmark &  &  & \checkmark \\
Reference signal $r_k$                 & \checkmark & \checkmark & \checkmark & \checkmark & \checkmark \\
Indoor temperature $T_k^i$             & \checkmark & \checkmark & \checkmark & \checkmark & \checkmark \\
Outdoor temperature $T_k^{\text{out}}$ & \checkmark & \checkmark & \checkmark & \checkmark & \checkmark \\
BESS SOC $\text{SOC}_k^i$           & \checkmark & \checkmark & \checkmark & \checkmark & \checkmark \\
Baseline load $y_k^{\text{base}}$      &  & \checkmark &  &  & \checkmark \\
Thermal model $(a_i,b_i,c_i,d_i)$      &  & \checkmark &  &  & \checkmark \\
HVAC power $P_i^{\text{HVAC}}$         &  & \checkmark &  &  & \checkmark \\
Local observation $o_k^i$              &  &  & \checkmark & \checkmark & \checkmark \\
District state $s_k$                  &  & \checkmark &  &  & \checkmark \\
\bottomrule
\end{tabular}}
\end{table}

\begin{table*}[t]
\centering
\caption{Coordination architectures and control responsibilities for the five controllers evaluated. CTDE = centralized training with decentralized execution.}
\label{tab:coordination}
\begin{tabular}{llll}
\toprule
Controller & Architecture & Observations used & Decision maker (assets) \\
\midrule
RBC
& Decentralized
& --
& Local agent $i$ (HVAC, BESS) \\
MPC
& Centralized
& $\{T_k^i,\;\text{SOC}_k^i,\;\text{constraints}\}_{i=1}^N$
& Central agent (HVAC, BESS) \\
I-SAC
& Decentralized
& Local $o_k^i$, reference signal $r_k$
& Local agent $i$ (HVAC, BESS) \\
MAPPO
& CTDE
& Local $o_k^i$, reference signal $r_k$
& Local agent $i$ (HVAC, BESS) \\

Hybrid (MPC+SAC)
& Hierarchical
& MPC: aggregated load, reference signal 
& Central agent (BESS) 
 \\
& & SAC: local $o_k^i$ & Local agent $i$ (HVAC)\\
\bottomrule
\end{tabular}
\end{table*}

\subsubsection{Rule-based controller (RBC)}

The rule-based controller serves as a transparent operational baseline that enforces thermal comfort while performing time-dependent BESS scheduling at the building level. Each building follows a fixed \emph{time-of-use} (TOU) policy: the BESS is charged during off-peak hours (22:00--08:00) and discharged during peak hours (14:00--21:00) to reduce net electricity demand. Heating is regulated locally using a hysteresis thermostat to maintain indoor temperature within a predefined comfort band. The controller requires no training and relies only on current measurements.

\subsubsection{Centralized Model Predictive Control (MPC)}

A centralized convex MPC controller jointly optimizes HVAC and BESS operation for all buildings over a receding horizon $H$, following standard formulations for buildings~\cite{drgona_all_2020}. District demand is modeled as
\begin{equation}
y_k
=
y_k^{\text{base}}
+
\sum_{i=1}^N p_{i,k}^{\text{batt}}
+
\sum_{i=1}^N P_i^{\text{HVAC}}\,u_{i,k}.
\end{equation}

and indoor temperature dynamics are approximated by first-order linear models,
\begin{equation}
T_{k+1}^i = a_i T_k^i + b_i T_k^{\text{out}} + c_i P_i^{\text{HVAC}} u_{i,k} + d_i ,
\end{equation}

The BESS state-of-charge (SOC) dynamics are governed by
\begin{equation}
\mathrm{SOC}_{k+1}^i
=
\mathrm{SOC}_k^i
+
\frac{p_{i,k}^{\text{batt}}}{E_{\text{cap}}^i}.
\end{equation}

At each time step, the controller solves the quadratic program
\begin{equation}
\min_{\{u_{i,k},\,p_{i,k}^{\text{batt}}\}}
\sum_{k=t}^{t+H-1}
\Big(
w_{\text{track}}\lVert y_k-r_k\rVert^2
+
w_{\text{comfort}}\!\sum_{i=1}^N v_k^i
+
w_{\text{ctrl}}\!\sum_{i=1}^N \lVert u_{i,k}\rVert^2
\Big).
\end{equation}

subject to thermal comfort, BESS state-of-charge, and power constraints, and applies the first control action.

\begin{equation}
T_i^{\min} - s_{k,\mathrm{lo}}^i \le T_k^i \le T_i^{\max} + s_{k,\mathrm{hi}}^i,
\quad s_{k,\mathrm{lo}}^i, s_{k,\mathrm{hi}}^i \ge 0,
\end{equation}
\begin{equation}
0 \le u_{i,k} \le 1,
\quad
p_i^{\min} \le p_{i,k}^{\text{batt}} \le p_i^{\max},
\quad
\mathrm{SOC}_i^{\min} \le \mathrm{SOC}_{i,k} \le \mathrm{SOC}_i^{\max}.
\end{equation}

Following standard MPC practice, thermal comfort is enforced as a soft constraint using slack variables to ensure numerical feasibility during periods when comfort violations are unavoidable~\cite{drgona_all_2020}.

The weights $w_{\text{track}}$, $w_{\text{comfort}}$, and $w_{\text{ctrl}}$ balance district-level tracking accuracy, thermal comfort preservation, and HVAC control effort, respectively. Perfect forecasts of baseline load, outdoor temperature, and the reference signal are assumed to provide an upper-bound benchmark for centralized control.

\subsubsection{Independent Soft Actor--Critic (I-SAC)}

We study I-SAC, where each building $i$ learns an individual stochastic policy
\begin{equation}
\pi_\theta^i(a_k^i \mid o_k^i),
\end{equation}
using only its local observation $o_k^i$ and the broadcast reference signal $r_k$, while receiving a shared global reward \eqref{eq:globalreward}. This follows the independent learning paradigm in cooperative multi-agent reinforcement learning \cite{lowe_multi-agent_2017} and is based on the SAC algorithm~\cite{haarnoja_soft_2018}.

\paragraph{Reward design}

All agents are trained using a shared district-level reward to promote cooperative behavior. The tracking error is defined as
\begin{equation}
e_k = y_k - r_k .
\end{equation}

A Huber loss is used for robustness:
\begin{equation}
\mathcal{L}_{\text{track}}(e_k) =
\begin{cases}
\frac{1}{2} e_k^2, & |e_k| \le \delta, \\
\delta(|e_k| - \frac{1}{2}\delta), & \text{otherwise}.
\end{cases}
\end{equation}

Thermal comfort violations are penalized by
\begin{equation}
\mathcal{L}_{\text{comfort}} =
\frac{1}{N}\sum_{i=1}^N
\Big[
\max(0, T_k^i - T_{\max})
+
\max(0, T_{\min} - T_k^i)
\Big].
\end{equation}

The global reward is
\begin{equation}
R_k =
-
\Big(
w_{\text{track}} \mathcal{L}_{\text{track}}
+
w_{\text{comfort}} \mathcal{L}_{\text{comfort}}
\Big),
\label{eq:globalreward}
\end{equation}
and is broadcast to all agents.

\paragraph{Policy optimization}

Each agent maximizes the entropy-regularized objective
\begin{equation}
J_{\pi}^i =
\mathbb{E}\!\left[
Q_\phi^i(o_k^i,a_k^i)
-
\alpha \log \pi_\theta^i(a_k^i \mid o_k^i)
\right],
\end{equation}
while the critic minimizes
\begin{equation}
J_Q^i =
\mathbb{E}\!\left[
\left(
Q_\phi^i(o_k^i,a_k^i)
-
(R_k+\gamma V_{\bar{\phi}}^i(o_{k+1}^i))
\right)^2
\right].
\end{equation}

\subsubsection{Multi-Agent Proximal Policy Optimization (MAPPO)}

We employ MAPPO under the centralized-training–decentralized-execution (CTDE) paradigm \cite{yu_surprising_2022}. A centralized critic estimates the state value
\begin{equation}
V(s_k).
\end{equation}

Training uses the global reward $R_k$ from~\eqref{eq:globalreward}. The temporal-difference residual is
\begin{equation}
\delta_k = R_k + \gamma V(s_{k+1}) - V(s_k),
\end{equation}

and the generalized advantage estimate is
\begin{equation}
A_k = \sum_{j=0}^{\infty} (\gamma\lambda)^j \delta_{k+j},
\end{equation}
where $\gamma$ is the discount factor and $\lambda\in[0,1]$ is the GAE smoothing parameter.

Policy updates follow the clipped PPO objective
\begin{equation}
J_{\text{PPO}} =
\mathbb{E}\!\left[
\min\!\left(
\rho_k^i A_k,
\text{clip}(\rho_k^i,1-\epsilon,1+\epsilon) A_k
\right)
\right],
\end{equation}
where $\rho_k^i$ is the probability ratio between the updated and previous policies of agent $i$.
At execution time, each building acts using only its local observation, ensuring privacy preservation and scalability.

\subsubsection{Hybrid MPC--SAC}

The hybrid controller decomposes decision making across time and system scales. A centralized MPC optimizes building-level BESS charging and discharging to enforce district-level tracking of the reference signal, while decentralized SAC agents independently regulate building-level HVAC to preserve thermal comfort.
At each time step,
\begin{align}
a_{k,\text{hvac}}^i &= \pi_\theta^i(o_k^i), \\
p_{i,k}^{\text{batt}} &= \arg\min J_{\text{MPC}},
\end{align}
where $\pi_\theta^i$ denotes the learned HVAC policy of building $i$ and $p_{i,k}^{\text{batt}}$ is the battery power setpoint computed by the centralized MPC.

This architecture combines the constraint-handling and coordination capabilities of MPC with the adaptability and privacy-preserving execution of decentralized SAC, centralizing coordination where it is most effective and decentralizing control where local adaptation is critical.

\subsection{Performance Metrics}
\label{sec:metrics}

Controller performance is evaluated using metrics that quantify (i) district-level tracking accuracy, (ii) thermal comfort preservation, and (iii) spatial distribution of control actions.

\subsubsection{District load tracking accuracy}

These metrics evaluate how accurately the aggregated district load follows the reference signal, capturing both systematic bias and short-term variability.

\paragraph{Normalized Mean Bias Error (NMBE)}
NMBE measures the average signed tracking error relative to the mean reference demand, indicating persistent over- or under-consumption.

\begin{equation}
\mathrm{NMBE} =
\frac{100}{\overline{r}}
\cdot
\frac{1}{K}
\sum_{k=1}^{K}
(y_k - r_k),
\end{equation}
with $
\overline{r} = \frac{1}{K}\sum_{k=1}^{K} r_k
$.

\paragraph{Coefficient of Variation of RMSE (CVRMSE)}
CVRMSE quantifies the magnitude of tracking fluctuations around the reference, normalized by the average reference demand.

\begin{equation}
\mathrm{CVRMSE} =
\frac{100}{\overline{r}}
\sqrt{
\frac{1}{K}
\sum_{k=1}^{K}
(y_k - r_k)^2
}.
\end{equation}

\subsubsection{Thermal comfort}

These metrics quantify frequency and magnitude of indoor temperatures deviations from an acceptable comfort range. The instantaneous comfort violation is
\begin{equation}
v_k^i =
\max(0, T_k^i - T_{\max})
+
\max(0, T_{\min} - T_k^i).
\end{equation}
Where $T_{\min}$ and $T_{\max}$ are the temperatures defining the thermal comfort band. Thus, the total \emph{Exceedance hours} are given by
\begin{equation}
H_i =
\sum_{k=1}^{K}
\mathbb{I}(v_k^i > 0).
\label{eq:excedance_hours}
\end{equation}
where $\mathbb{I}$ is the indicator function equal to 1 if its argument is positive, and 0 else. \eqref{eq:excedance_hours} can also be expressed as a percentage, i.e., $P_i =
\frac{100}{K}H_i$.

At the neighborhood scale, the mean exceedance percentage is:
\begin{equation}
\overline{P} =
\frac{1}{N}
\sum_{i=1}^{N} P_i .
\end{equation}

Magnitude and duration of comfort violations are assessed by 
\begin{equation}
K_i =
\sum_{k=1}^{K} v_k^i \cdot \Delta t , \quad [\text{K}\cdot\text{h}]
\end{equation}
and its mean across all buildings is $ \overline{K} =
\frac{1}{N}\sum_{i=1}^{N} K_i .
$

\subsubsection{Spatial control variability}

This evaluates how evenly control actions are distributed across buildings, i.e., whether flexibility is concentrated on a small subset of buildings (high variability) or shared uniformly (low variability). For controller $c$ and building $i$, the spatial variability at time $k$ is
\begin{equation}
\sigma(\Delta y(k)) =
\sqrt{
\frac{1}{N}
\sum_{i=1}^{N}
\left(
\Delta y_i(k) - \overline{\Delta y(k)}
\right)^2
},
\label{eq:sigma}
\end{equation}
where $
\Delta y_i(k) = y_{i,k}^{c} - y_{i,k}^{\mathrm{RBC}}$ is the deviation from the RBC baseline, 
and 
\begin{equation}
\overline{\Delta y(k)} = \frac{1}{N}\sum_{i=1}^{N} \Delta y_i(k).
\end{equation}

To summarize behavior over the evaluation horizon, we report the median spatial variability:
\begin{equation}
\mathrm{SV}_{\mathrm{med}} = \mathrm{median}_{k}\!\left[\sigma(\Delta y(k))\right].
\end{equation}

\subsection{Experimental Setup}

All controllers are evaluated in the CityLearn environment \cite{nweye_citylearn_2025} on a district of $N=25$ heterogeneous residential buildings equipped with a heat pump, rooftop photovoltaics (PV), and a battery energy storage system (BESS), as summarized in Table~\ref{tab:table_buildings}. The experiments are conducted under a cold-climate setting representative of Vermont. Indoor comfort bounds are defined between $T_{\min} =20^\circ$ C and $T_{\max} =24^\circ$C. The installed BESS capacities across buildings range from \textit{8.0 to 24.5 kWh}. Rooftop PV systems exhibit peak hourly generation between approximately \textit{2 and 14 kWh} depending on building size and orientation. The baseline mean hourly net electricity demand per building ranges from approximately \textit{1.5 to 8 kWh}.

The SAC, MAPPO, and the SAC component of the hybrid controller are trained using a 30-day (January) simulation horizon with hourly control resolution and evaluated on an unseen test period corresponding to 28-day February. During testing, all learned policies are frozen and no further parameter updates are performed. The centralized MPC controller uses training data (January) to identify building-level thermal models, which are then kept fixed during test evaluation. All quantitative results reported in this paper correspond to the test period, unless otherwise noted.

Table~\ref{tab:hyperparameters} in the Appendix summarizes the hyperparameters used in the experiments. The full codebase and processed datasets  are available online \footnote{https://github.com/intelligent-environments-lab/buildsys-26-coordination}
.

\begin{table*}[tbh]
\centering
\caption{District-level tracking accuracy, thermal comfort, and spatial variability for all controllers during training (January) and test (February). Exceed. = comfort exceedance; K$\cdot$h = cumulative discomfort; $\mathrm{SV}_{\mathrm{med}}$  = median spatial variability wrt to RBC. Bold values indicate best performance per column.}
\label{tab:results_tracking_comfort}

\setlength{\tabcolsep}{6pt}

\begin{tabular}{lrrrrrrr}
\toprule
\multirow{2}{*}{Controller} &
\multicolumn{2}{c}{Tracking (Train)} &
\multicolumn{2}{c}{Tracking (Test)} &
\multicolumn{2}{c}{Comfort (Test)} &
Spatial Variability (Test) \\
\cmidrule(lr){2-3}\cmidrule(lr){4-5}\cmidrule(lr){6-7}\cmidrule(lr){8-8}
& NMBE [\%]& CVRMSE [\%]& NMBE [\%]& CVRMSE [\%]& Exceed. [\%] & K$\cdot$h & $\mathrm{SV}_{\mathrm{med}}$ [kWh]\\
\midrule

RBC & 37.93 & 71.45 & 42.32 & 74.62 & \textbf{14.09} &\textbf{26.82} & -- \\

MPC &\textbf{1.59}& \textbf{38.59} & 8.81 & \textbf{56.70} & 24.78 & 105.91 & 1.85 \\

SAC & 23.27 & 57.87 & 24.79 & 61.48 & 17.54 & 46.90 & 1.70 \\

MAPPO & 33.11 & 78.06 & 37.76 & 71.73 & 19.17 & 125.58 & 1.19 \\

Hybrid (MPC+SAC) & 2.31 & 88.05 & \textbf{4.80} & 57.09 & 16.84 & 102.50 & \textbf{1.18} \\

\bottomrule
\end{tabular}
\end{table*}

\begin{figure*}[tbh]
    \centering
    \includegraphics[trim=0.25cm 0.6cm 0.25cm 0.25cm,clip,width=\textwidth]{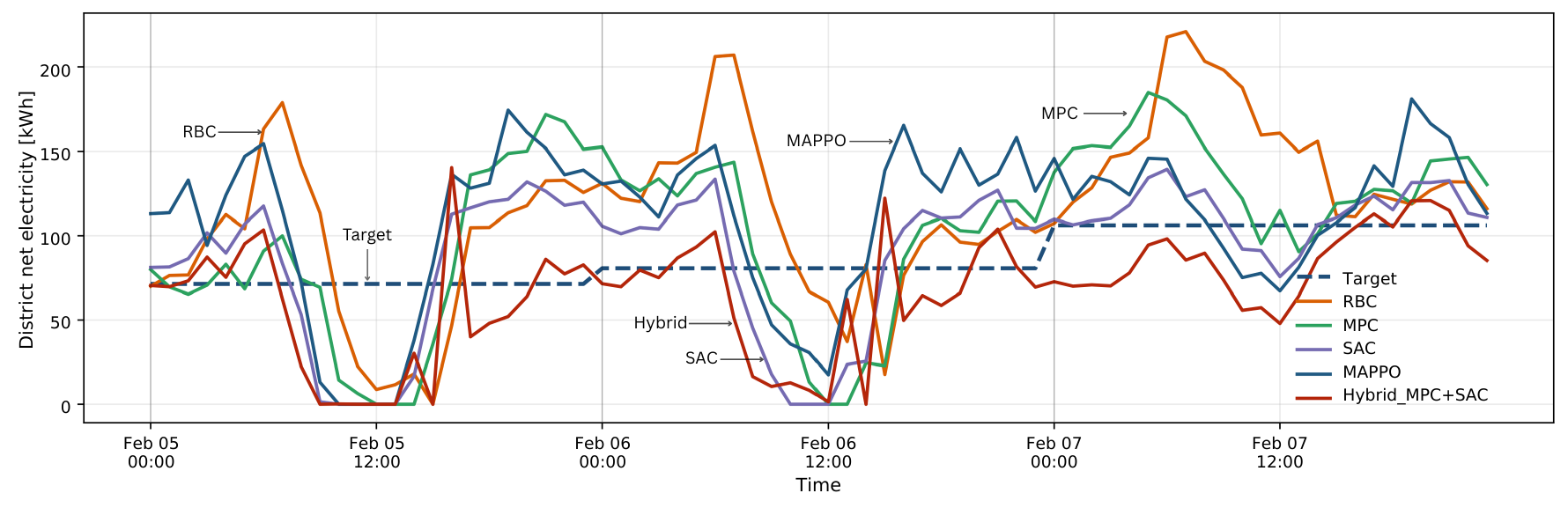}
    \caption{District net electricity load during a representative 3-day test period (February). The dashed line shows the reference signal and solid lines correspond to RBC, MPC, SAC, MAPPO, and the hybrid MPC+SAC controller.}

    \label{fig:district_comparison}
\end{figure*}

\section{Results}
\label{s:results}

Table~\ref{tab:results_tracking_comfort} summarizes district-level tracking accuracy, thermal comfort, and spatial control variability performance for all controllers on the training (January) and test (February) periods. We discuss them now in the following.

\subsection{District Scale Analysis}
Rule-based control (RBC) exhibits large systematic tracking errors, with test NMBE exceeding 40\% and CVRMSE above 70\%, confirming its inability to follow the reference signal under dynamic operating conditions. The hybrid MPC+SAC controller achieves the lowest tracking bias on the test set (NMBE = 4.80\%), followed by MPC (8.81\%). I-SAC and MAPPO show higher bias and variability, indicating weaker robustness to disturbance uncertainty. The centralized MPC controller shows a notable train--test performance gap, with considerably lower tracking error during training than testing. This degradation reflects model mismatch under unseen operating conditions and disturbance uncertainty, as thermal and load models identified from January data are applied unchanged during the February test period.

Figure~\ref{fig:district_comparison} shows district-level trajectories during the test period.
The hybrid controller follows the reference signal across both peak and ramping periods the closest, while SAC and MAPPO exhibit delayed responses during rapid load changes. Importantly, low average tracking bias (NMBE) does not necessarily imply accurate hourly tracking. In particular, MPC achieves relatively small bias but exhibits a high CVRMSE (56.70\%), indicating pronounced temporal fluctuations around the reference.

\begin{figure}[t]
    \centering
    \includegraphics[trim=0.25cm 0.25cm 0.25cm 0.25cm,clip,width=\columnwidth]{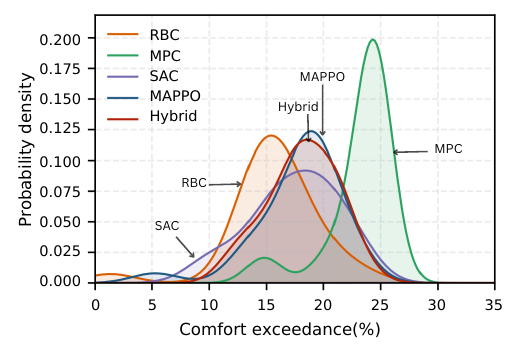}
    \caption{Probability density of building-level comfort exceedance across all 25 buildings.}
    \label{fig:comfort_exceedance_dist}
\end{figure}

Figure~\ref{fig:comfort_exceedance_dist} shows the distribution of building-level comfort exceedance for all controllers. The distributions highlight clear differences in how control strategies impact occupant thermal conditions across the building population.
Comparing Tab.~\ref{tab:results_tracking_comfort} and Figure~\ref{fig:comfort_exceedance_dist}, we can see a clear trade-off between district-level tracking accuracy and occupant comfort: While centralized MPC reduces tracking error relative to RBC, it incurs substantially higher comfort exceedance (24.78\%) and cumulative discomfort (105.9~K$\cdot$h), reflecting aggressive control to satisfy district-level objectives.
Hybrid and SAC achieve the lowest comfort violation rates among the advanced methods (16.84\% and 17.54\%, respectively), approaching the performance of RBC (14.09\%).
Moreover, SAC exhibits substantially lower cumulative discomfort (46.9~K$\cdot$h), indicating that it limits not only the frequency but also the severity of temperature deviations.
In contrast, MAPPO shows the highest cumulative discomfort (125.6~K$\cdot$h) and a broader exceedance distribution, indicating large and persistent comfort violations for a subset of buildings despite improvements in tracking bias.

\subsection{Building Scale Analysis}
\begin{figure*}[t]
    \centering
    \includegraphics[trim=0.25cm 0.25cm 0.25cm 0.25cm,clip,width=\textwidth]{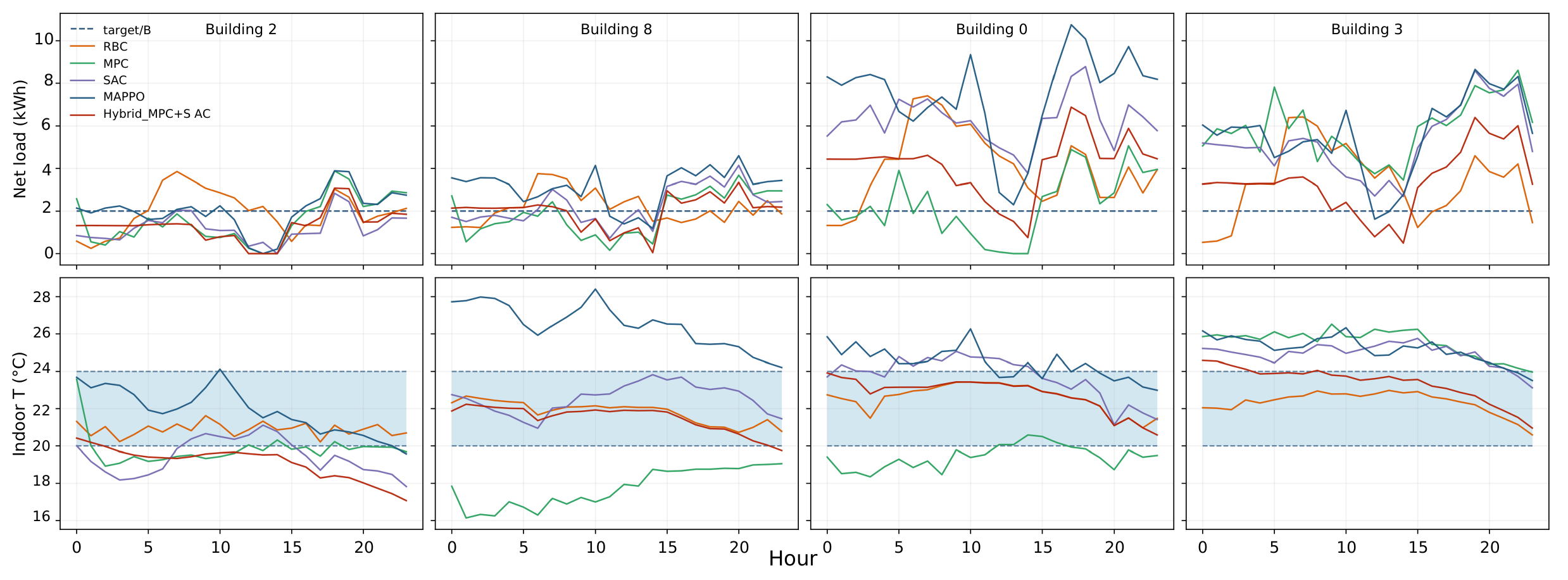}
    \caption{Net electricity consumption (top) and indoor temperature (bottom) for four representative buildings during a 24-hour test period. Shaded regions indicate the comfort bounds. Buildings B0 and B3 are higher-demand; B2 and B8 are lower-demand.}

    \label{fig:per_building_diagnostics}
\end{figure*}

We investigate how the aggregated load in Figure~\ref{fig:district_comparison} is reflected at the building level. In Figure~\ref{fig:per_building_diagnostics} we show net electricity demand and indoor temperature for four representative buildings, two with lower-demand (B2, B8) and two with higher-demand (B0, B3). MPC drives indoor temperatures toward the lower comfort boundary in some buildings, exploiting their thermal flexibility to achieve tracking, while MAPPO, in contrast, exhibits larger deviations from the comfort band in several buildings and less consistent tracking of the district reference. In contrast, SAC and hybrid controller  maintain temperatures closer to the comfort band, limiting long comfort violations while still contributing to system-level regulation.

Figure~\ref{fig:per_building_diagnostics} reveals that different controllers concentrate tracking effort on different buildings, with corresponding comfort consequences. In B0, MPC exhibits aggressive load suppression that drives indoor temperature below the comfort bound, while the hybrid controller achieves similar tracking with shallower violations. In B3, SAC causes prolonged overheating, whereas the hybrid maintains temperatures closer to the comfort range. The lower-demand buildings show heterogeneous patterns: MPC induces persistent under-heating in B8, while both SAC and the hybrid experience temporary violations in B2. SAC and MAPPO exhibit weaker and more delayed load responses than MPC or the hybrid controller.

\begin{figure}[tb]
    \centering
    \includegraphics[trim=0.25cm 0.25cm 0.25cm 0.25cm,clip,width=\columnwidth]{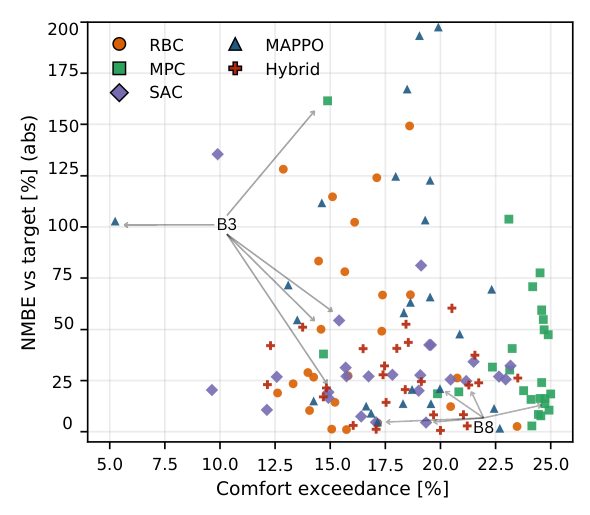}
    \caption{Trade-off between load tracking and thermal comfort. Each point represents one building under one controller (N=25 buildings × 5 controllers). Lower values indicate better performance on each axis. Buildings B3 and B8 are annotated for reference.}
    \label{fig:comfort_nmbe}
\end{figure}

These results indicate that neither tracking contributions nor comfort violations are uniformly distributed across buildings. We examine this trade-off further in Figure~\ref{fig:comfort_nmbe}, where we plot building-level comfort exceedance against absolute tracking error for each building/controller combination. Tighter clusters for each controller (marker/color) indicate similar performance trade-off for each building, whereas large clusters indicate dissimilar performance.  Controllers achieving lower tracking error (bottom of the plot) generally incur higher comfort penalties (right of the plot), while methods prioritizing comfort preservation exhibit larger deviations from the district reference. The hybrid MPC+SAC controller occupies a favorable intermediate regime, combining low tracking bias with limited comfort degradation.

Figure~\ref{fig:comfort_nmbe} also highlights strong building-dependent controller performance. For example, B3 appears in markedly different regions of the trade-off space across controllers. Under MPC, it exhibits comfort exceedance close to RBC but with large tracking error, while SAC shows similar behavior to RBC. The hybrid controller noticeably reduces tracking error with the same comfort violations, whereas MAPPO improves comfort considerably but incurs larger tracking errors (still better than MPC though). B8 experiences substantial comfort degradation and increased tracking error under most controllers compared to RBC, which remains close to the reference while maintaining the lowest comfort exceedance.

\begin{figure}[tb]
    \centering
    \includegraphics[trim=0.25cm 0.25cm 0.25cm 0.25cm,clip,width=\columnwidth]{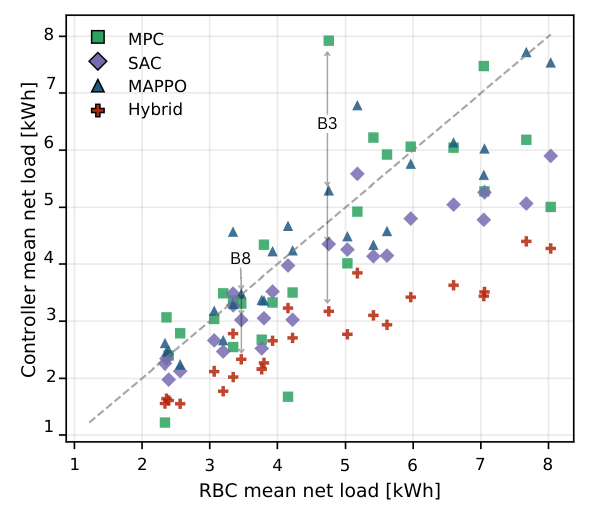}
    \caption{Mean net electricity load per building under each controller vs RBC baseline. Points on the diagonal indicate no change from baseline; points below (above) indicate load reduction (increase). Hybrid systematically reduces loads across buildings, while MPC shows high variance with selective exploitation of high-load buildings.}
    \label{fig:net_load_per_building}
\end{figure}

Figure~\ref{fig:net_load_per_building} provides a complementary perspective by comparing the mean net load of each building ($y$-axis) under different controllers against the RBC baseline ($x$-axis): Points below the diagonal indicate load reduction relative to RBC.
The hybrid controller consistently shifts buildings demand downward with a tight spread, indicating load reduction across all buildings in which higher-baseline load buildings contribute more strongly to load reduction. SAC also induces a reduction, but with a smoother, load-dependent trend.
In contrast, centralized MPC exhibits a much wider dispersion, suggesting selective and uneven utilization of flexibility resources, while MAPPO shows a scattered pattern. Only MPC and MAPPO increase the load, with one SAC building being the exception.
This heterogeneity is particularly evident in buildings B3 and B8. Under MPC, B3 exhibits a substantial load increase, while B8 experiences a small net load reduction relative to baseline. Under the hybrid controller, both buildings shift below the diagonal, contributing to district-level regulation in a more balanced manner. 

\begin{figure}[tb]
    \centering
    \includegraphics[trim=0.15cm 0.25cm 0.25cm 0.1cm,clip, width=\columnwidth]{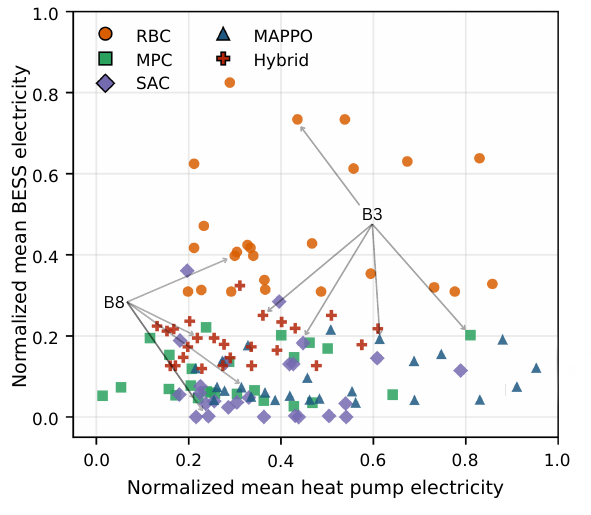}
    \caption{Building-level relationship between normalized mean heat pump and battery electricity usage under different controllers. Each point represents one building under one controller.}
    \label{fig:batt_hp_scatter}
\end{figure}

Figure~\ref{fig:batt_hp_scatter} shows how different coordination strategies exploit the two flexibility assets (heat pump and BESS). Larger groups for a controller (same color/marker) represent heterogeneous controller behavior across buildings, while tighter groups correspond to similar behavior in the buildings. RBC activates both HVAC and BESS the strongest in many buildings. Centralized MPC, SAC and MAPPO rely primarily on HVAC modulation with limited battery engagement only in few buildings, leading to uneven exploitation of assets. In contrast the hybrid MPC+SAC controller achieves the most balanced joint use of thermal and electrical storage across buildings while also distributing the actions similarly across the buildings (smallest group). The behavior of B3 and B8 further highlights how the hybrid controller redistributes flexibility more evenly, avoiding the extreme asset usage observed under RBC and MPC.

\subsection{Spatial Variability}
\begin{figure}[tb]
    \centering
    \includegraphics[trim=0.25cm 0.25cm 0.25cm 0.25cm,clip,width=\columnwidth]{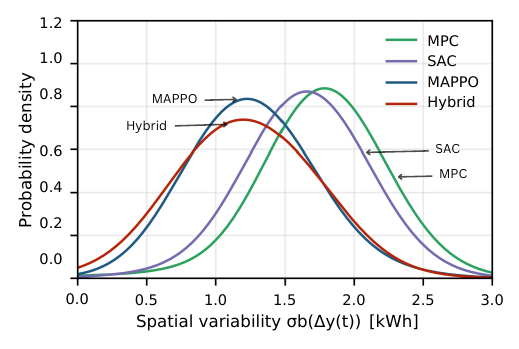}
    \caption{
    Distribution of hourly spatial control variability, $\sigma_b(\Delta y(t))$ (see~\eqref{eq:sigma}). Lower values indicate more uniform distribution of control actions across the building population.}
    \label{fig:variability_distribution}
\end{figure}

Our findings in the previous subsection
(Figs.~\ref{fig:per_building_diagnostics}--\ref{fig:batt_hp_scatter}) suggest that controller performance trade-offs as well as asset activation are different for each building. We now further investigate this using the spatial variability concept (see~\eqref{eq:sigma}). Figure~\ref{fig:variability_distribution} presents the distribution of temporal spatial control variability across buildings. MAPPO and the hybrid MPC+SAC controller show comparatively low median variability (1.19 and 1.18~kWh), indicating more uniform allocation of control actions across buildings. SAC shows moderately higher dispersion (1.70~kWh), while MPC exhibits the largest variability with median 1.85~kWh, corresponding to sporadic but pronounced spatial imbalances. 

\begin{figure*}[tb]
    \centering
    \includegraphics[trim=0.25cm 0.65cm 0.25cm 0.25cm,clip,width=\textwidth]{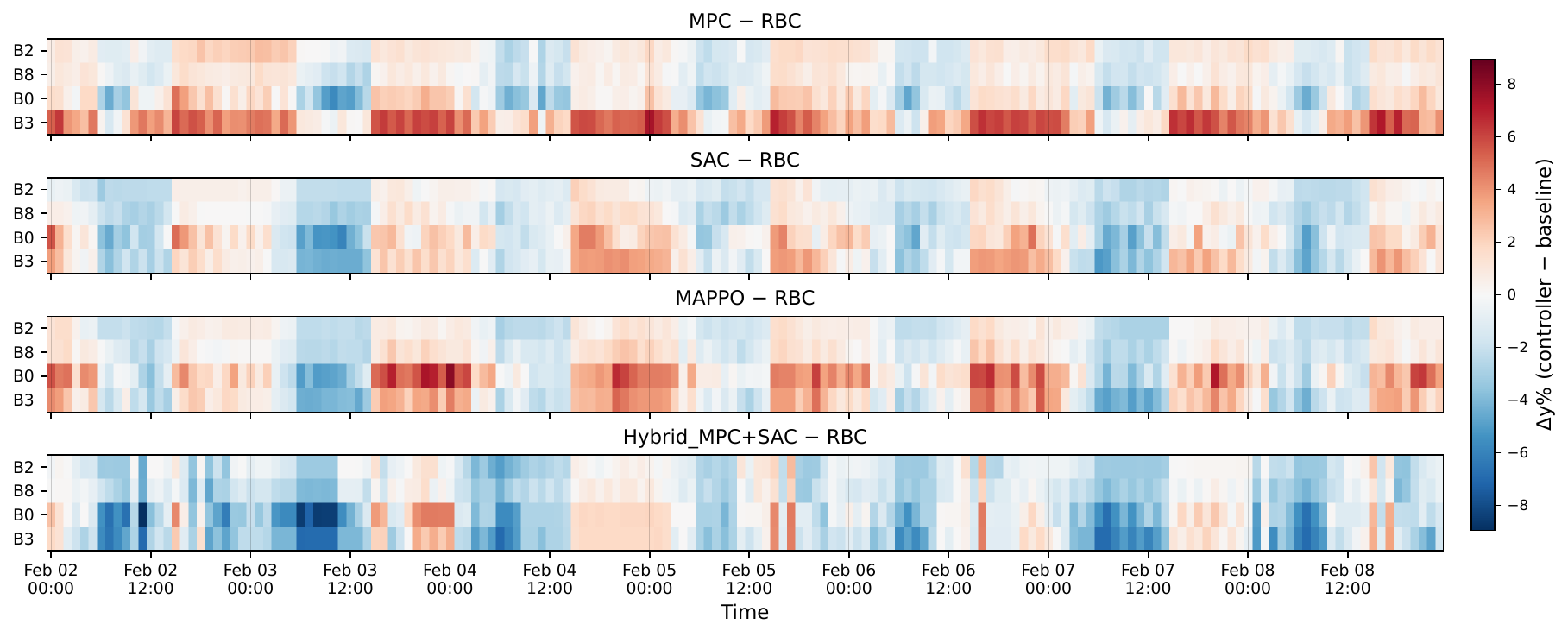}
    \caption{Building-level changes in net electricity consumption relative to RBC, $\Delta y_b(k)$ for buildings B0, B2, B3, and B8 during February 2–8. Color scale indicates load increase (red) or decrease (blue) in kWh. Each row corresponds to one controller}
    \label{fig:delta_heatmaps}
\end{figure*}

These patterns also hold on an hourly basis as shown in the building-level heatmaps in Figure~\ref{fig:delta_heatmaps} for the four representative buildings (B0/2/3/8). Centralized MPC concentrates large corrective actions on a small subset of buildings, most notably higher-demand buildings, e.g., B0 and B3, leading to repeated exploitation of the same units. In contrast, SAC and the hybrid controller distribute adjustments more evenly across buildings, avoiding persistent reliance on individual assets. MAPPO occupies an intermediate regime; and while it does not consistently target the same buildings, it exhibits intermittent building-specific bursts, particularly for B0, resulting in occasional spatial imbalance. 

\begin{figure}[t]
    \centering
    \includegraphics[trim=0.25cm 0.6cm 0.25cm 0.25cm,clip, width=\columnwidth]{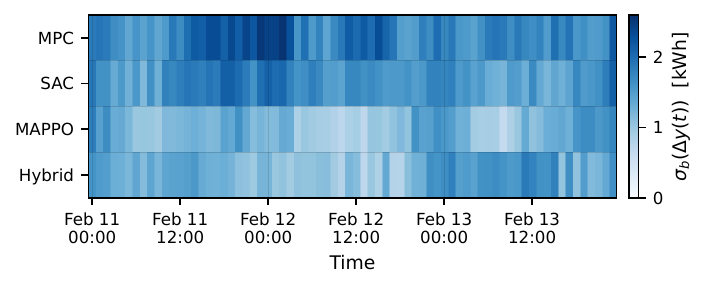}
    \caption{Hourly spatial control variability measured as $\sigma_b(\Delta y(k))$ across buildings for each controller. Higher values (darker colors) indicate greater dispersion of control actions across buildings at that hour.}
    \label{fig:hourly_variability_heatmap}
\end{figure}

Finally, Figure~\ref{fig:hourly_variability_heatmap} highlights the temporal structure of these effects. MPC displays recurring peaks in spatial variability during certain periods, indicating coordination breakdowns under system stress. SAC and the hybrid controller maintain smoother coordination patterns over time, with consistently lower across-building dispersion. MAPPO again lies between these extremes, exhibiting generally moderate variability punctuated by episodic spikes during dynamic operating conditions.

\section{Discussion}
\label{s:discussion}
Our results demonstrate that coordination architecture determines the trade-off between district-level tracking and building-level thermal comfort through two primary mechanisms: (i) the degree of centralization in decision-making, and (ii) the coupling between assets (HVAC vs.\ BESS) and objectives (tracking vs.\ comfort). Centralized MPC achieves the lowest tracking bias by jointly optimizing all assets toward the district objective, confirming prior findings~\cite{el_geneidy_contracted_2020,lefebure_distributed_2022}, but this global view leads it to exploit thermally flexible buildings disproportionately, concentrating comfort violations on a subset of the population. Fully decentralized approaches (SAC, MAPPO) distribute control effort more evenly but lack the coordination capacity to sustain accurate tracking under continuous grid-following requirements~\cite{zhang_multi-agent_2023,savino_scalable_2025}. The hybrid MPC+SAC decouples the problem: centralized optimization handles district-level coordination through BESS activation, while decentralized learning preserves building-level comfort autonomy. This structural decomposition explains its favorable performance and aligns with the CityLearn Challenge 2022 results, where all top-performing solutions among 600+ participants employed hybrid architectures, while pure RL approaches underperformed~\cite{nweye_citylearn_2022}.

The finding that RBC achieves the best comfort preservation (14.09\% exceedance, 26.8~K$\cdot$h) while failing at tracking (42\% NMBE) quantifies the value proposition of advanced control: not comfort improvement, but the ability to deliver grid services \emph{while} maintaining acceptable comfort. This reframes the evaluation of demand response controllers: comfort should be an objective to achieve rather than a soft constraint to satisfy.

Although spatial variability is evaluated here as a technical metric, it carries practical implications for demand response program design. Centralized MPC's tendency to repeatedly exploit the same thermally flexible buildings raises equity concerns regarding the distribution of comfort impacts: some occupants bear disproportionate comfort costs while others are largely unaffected. This connects to broader discussions of equity in local energy systems, where uneven allocation of costs and benefits can undermine participant engagement and long-term program viability~\cite{soares_review_2024}. The hybrid architecture's low spatial variability (SV$_\text{med}$ = 1.18~kWh vs.\ MPC's 1.85~kWh) suggests that hierarchical decomposition may offer not only technical advantages but also more equitable burden-sharing across building occupants.

Most demand response applications target coarse load reduction objectives such as peak shaving. In contrast, we address continuous tracking of a constant daily reference signal, which is a deliberately challenging formulation that exposes coordination effects potentially hidden under smoother objectives. We hypothesize that relaxing this constraint (e.g., allowing time-varying targets or event-based signals) would reduce comfort and coordination trade-offs, but would not reveal the structural differences between architectures observed here. This highlights reference signal design as a critical but underexplored factor in coordinated control evaluation.

\paragraph{Limitations}
First, thermal comfort is evaluated using static temperature bands, which provide a conservative but incomplete proxy for real occupant comfort that is adaptive, heterogeneous, and influenced by behavioral and contextual factors. Second, the centralized MPC assumes perfect forecasts; real-world implementations would face forecast errors that likely degrade tracking performance relative to the upper-bound results reported here. Third, scalability and computational cost are not explicitly evaluated, and controller performance may change as district size increases or communication constraints become binding. Fourth, the comparison between MPC (which requires only model identification) and RL methods (which require 30 days of training data) may not be fair in data-limited deployment scenarios. Finally, results are reported for a single cold climate and a deliberately challenging constant reference signal; performance rankings may differ under alternative weather conditions, milder reference profiles, or event-based demand response objectives.

\paragraph{Future work}
Future research should extend this evaluation framework to dynamic and personalized comfort models that capture variability in thermal preferences~\cite{luo_effects_2022}. The robustness of coordination architectures across different reference signal designs, e.g., event-based curtailment, ramping services, and frequency regulation, remains unexplored. Finally, explicit equity constraints on the distribution of comfort impacts could be incorporated into the control formulation to guarantee equitable distribution of comfort impacts across buildings.

\section{Conclusion}
\label{s:conclusion}
As building districts transition from event-based demand response toward continuous flexibility provision, the choice of coordination architecture becomes a design decision with direct consequences for both grid performance and occupant outcomes. We show that centralized MPC delivers accurate tracking (8.8\% NMBE) but concentrates comfort violations on a subset of thermally flexible buildings (24.8\% exceedance); decentralized RL preserves comfort more evenly but cannot sustain tracking under continuous grid-following requirements. The hybrid MPC+SAC architecture resolves this tension by separating district-level BESS coordination from building-level HVAC regulation, achieving the lowest tracking bias (4.8\% NMBE) and moderate comfort impact (16.8\% exceedance) without concentrating control burden. These results suggest that effective coordination requires matching control scope to objective scale, and that the spatial distribution of comfort impacts should be a standard evaluation dimension for district-level demand response.

\begin{acks}
\label{s:acknowledgement}
\begin{anonsuppress}
This work has been carried out within the framework of the International Energy Agency (IEA) Energy in Buildings and Communities (EBC) Annex 96: “Grid Integrated Control of Buildings” (https://annex96.iea-ebc.org). The authors gratefully acknowledge the support of the IEA EBC Annex 96 research network and our collaboration partners. The research is supported by the Netherlands Enterprise Agency (project MOOI224004). 
\end{anonsuppress}
\end{acks}

\FloatBarrier
\appendix
\section{Appendix}
Tab.~\ref{tab:hyperparameters} shows the hyperparameters used for the controllers, while Tab.~\ref{tab:table_buildings} shows RESSTOCK ID and metadata for the studied buildings.
\begin{table}[hbp]
\centering
\caption{Controller hyper-parameters used in the experiments. Tuned parameters (grid search) are marked with *.}
\label{tab:hyperparameters}
\footnotesize
\setlength{\tabcolsep}{6pt}
\begin{tabular}{llc}
\toprule
Controller & Hyperparameter & Value \\
\midrule

\multirow{7}{*}{MPC}
& Horizon $H$* & 12 \\
& Tracking weight $w_{\text{track}}$* & 0.5 \\
& Slack weight $w_{\text{slack}}$* & 50 \\
& Comfort weight $w_{\text{comfort}}$* & 300 \\
& Solver & OSQP \\
& Max iterations & $4\times 10^{5}$ \\
& Tolerance ($\varepsilon_{\text{abs}}, \varepsilon_{\text{rel}}$) & $10^{-4}$ \\

\midrule

\multirow{6}{*}{I-SAC}
& Discount factor $\gamma$ & 0.99 \\
& Entropy temperature $\alpha$* & 0.2 \\
& Actor learning rate* & $3\times10^{-4}$ \\
& Critic learning rate* & $3\times10^{-4}$ \\
& Batch size & 256 \\
& Replay buffer size & $10^6$ \\

\midrule

\multirow{6}{*}{MAPPO}
& Discount factor $\gamma$ & 0.99 \\
& GAE parameter $\lambda$ & 0.95 \\
& PPO clip $\epsilon$ & 0.2 \\
& Learning rate* & $3\times10^{-4}$ \\
& Batch size & 1024 \\
& Epochs per update & 10 \\

\bottomrule
\end{tabular}
\end{table}

\begin{table}[t]
\centering
\caption{Building archetypes and flexibility assets in the Vermont district ($N=25$), listed in building index order (B0--B24). Bold rows indicate buildings highlighted in building-level analyses (Figures 4, 5, 6, 7, 9) Building characteristics are taken from ResStock/BuildStock metadata.}
\label{tab:table_buildings}
\scriptsize
\setlength{\tabcolsep}{6pt}
\renewcommand{\arraystretch}{1.05}
\resizebox{\columnwidth}{!}{%
\begin{tabular}{lrrr}
\toprule
Bldg & ResStock ID & Floor area (m$^2$) & BESS capacity (kWh) \\
\midrule
\textbf{B0}  & \textbf{112208} & \textbf{306.7} & \textbf{21.6} \\
B1  & 147002 & 202.2 & 10.5 \\
\textbf{B2}  & \textbf{223581} & \textbf{157.0} & \textbf{10.8} \\
\textbf{B3}  & \textbf{199613} & \textbf{306.7} & \textbf{16.2} \\
B4  & 20199  & 306.7 & 20.0 \\
B5  & 216895 & 113.4 & 13.2 \\ 
B6  & 179247 & 157.0 & 16.0 \\
B7  & 245723 & 157.0 & 21.0 \\
\textbf{B8}  & \textbf{411001} & \textbf{157.0} & \textbf{13.5} \\
B9  & 319918 & 113.4 & 13.2 \\
B10 & 376570 & 202.2 & 16.2 \\
B11 & 408344 & 202.2 & 20.0 \\
B12 & 409896 & 247.3 & 10.8 \\
B13 & 247942 & 202.2 & 10.0 \\
B14 & 425540 & 202.2 & 14.0 \\
B15 & 4421   & 247.3 & 10.5 \\
B16 & 460412 & 306.7 &  8.0 \\
B17 & 467125 & 202.2 & 14.8 \\
B18 & 481052 & 202.2 & 20.0 \\
B19 & 485614 &  82.2 & 24.5 \\
B20 & 498771 & 157.0 & 10.5 \\
B21 & 525859 & 113.4 & 10.0 \\
B22 & 538628 & 157.0 & 10.5 \\
B23 & 76701  & 113.4 & 10.5 \\
B24 & 88386  & 113.4 & 20.0 \\
\bottomrule
\end{tabular}%
}
\end{table}

\clearpage
\bibliographystyle{ACM-Reference-Format}
\bibliography{References}

\end{document}